\documentclass[twocolumn,superscriptaddress,twocolumn,amsmath,amssymb,floatfix]{revtex4}
\usepackage{graphicx}

\begin{document}

\title{Large Variance and Fat Tail of Damage by Natural Disaster\footnote{http://ascelibrary.org/doi/abs/10.1061/9780784413609.277}}

\author{Hang-Hyun Jo}
\altaffiliation[Present address: ]{BK21plus Physics Division and Department of Physics, Pohang University of Science and Technology, Pohang 790-784, Republic of Korea}
\affiliation{Department of Biomedical Engineering and Computational Science, Aalto University, P.O. Box 12200, Espoo, Finland}
\author{Yu-li Ko}
\affiliation{Department of Economics, Rensselaer Polytechnic Institute, Troy, NY 12180, USA}

\date{\today}

\begin{abstract}
In order to account for large variance and fat tail of damage by natural disaster, we study a simple model by combining distributions of disaster and population/property with their spatial correlation. We assume fat-tailed or power-law distributions for disaster and population/property exposed to the disaster, and a constant vulnerability for exposed population/property. Our model suggests that the fat tail property of damage can be determined either by that of disaster or by those of population/property depending on which tail is fatter. It is also found that the spatial correlations of population/property can enhance or reduce the variance of damage depending on how fat the tails of population/property are. In case of tornadoes in the United States, we show that the damage does have fat tail property. Our results support that the standard cost-benefit analysis would not be reliable for social investment in vulnerability reduction and disaster prevention. 
\end{abstract}

\maketitle

\section{Introduction}

The 2011 Tohoku earthquake with magnitude of 9.0 or 9.1 struck the northeastern part of Japan. The earthquake was ranked as the fourth largest in the world. The overall cost of the damage has been estimated to be the tens of billions of US dollars, and 19,295 were killed and 359,073 houses were destroyed by the earthquake and resulting tsumani~\cite{Imamura2012}. This demonstrated that extreme events actually occur, implying that the probability of extreme events is small yet nonnegligible. This behavior can be characterized by large variances and fat tails in probability distributions both of damage and of natural disaster. Statistical properties of damage are also influenced by those of population/property exposed to disasters that have been described by large variances and fat tails. Despite its importance, the fat tail property and its implications in risk analysis have been far from being fully understood, although fat-tailed distributions of natural disaster and population/property have been intensively studied in statistical physics, geography, and other disciplines. 

In order to better understand the effect of natural disaster and population/property on damage, we devise a simple model by combining the occurrence distribution of natural disaster with population/property distributions. These distributions are assumed to have fat tails as various types of natural disasters, like earthquake and forest fire, and population/property are known to be described by fat-tailed or power-law distributions~\cite{Nishenko1995,Bak2002,Clauset2009}. We also take into account the tendency that population/property are spatially correlated partly due to the urbanization as there exist empirical and theoretical studies supporting such a tendency~\cite{Krugman1996}. We make additional assumptions. Firstly, the disasters are moving along a straight line. Secondly, the vulnerability is constant independent of the intensity of natural disaster and of exposed population/property. These assumptions can be easily relaxed to incorporate more generalized features, such as nonlinear dependence of vulnerability. Our model suggests large variances and fat tails of casualty and property damage by natural disaster.

We analytically solve the model for the limiting cases such that population/property are either fully uncorrelated or fully correlated in space. The more realistic, partially correlated cases are studied by numerical simulations because they are not analytically solvable. In general, the fat tail property of damage is expected to be affected by fat tail properties and spatial correlations of natural disaster and population/property. However, this is not always the case. We find that the fat tail of damage can be determined by either that of natural disaster or those of population/property, depending on which has a fatter tail than the other. The spatial correlations of population/property can enhance or reduce the fat tail property of damage, depending on how fat the tails of population/property distributions are. In order to empirically support our model, we analyze the dataset of casualties and property damages by tornadoes in the United States over 1970--2011. It is confirmed that the distributions of damage show fat tails.

Our research has implications in the effect of large variance and fat tail of damage on risk-related decision making. If we treat the variance of damage with the assumption of normal or thin-tailed distributions as in the typical risk analysis, it may lead to inefficient social investment to reduce vulnerability and consequently the damage. As Weitzman demonstrated with his dismal theorem, the fat tail property of uncertainty results in arbitrarily large or divergent expected loss, threatening the standard cost-benefit analysis~\cite{Weitzman2009,Nordhaus2011}. Our results emphasize the need to focus more on decision under uncertainty with fat-tailed distributions.

The paper is organized as following. Our model is introduced in Section Model and its analytic and numerical results are presented in Section Result, with empirical analysis of damages by tornadoes. In Section Conclusion, we conclude our paper with some remarks.

\section{Model}

\subsection{Background and assumptions}

Empirical findings about damages by natural disasters indicate that such damages can have large variances and be often characterized by fat-tailed or power-law distributions~\cite{Nishenko1995,Clauset2009}. The power-law distribution, e.g., for a damage $D$, is formally presented as
\begin{equation}
  \nonumber
  P(D)\sim D^{-\gamma},
\end{equation}
where $\gamma>0$ is a power-law exponent characterizing the degree of fat tail property. In such distributions, the statistics cannot be properly represented only by means due to very large or even diverging variances. The probability of extreme events is small yet nonnegligible, while the probability rapidly approaches zero for the thin-tailed cases such as exponential distributions.

In general, the risk or damage by natural disaster has been analyzed as a function of three components: natural disaster, population/property exposed to the disaster, and vulnerability of those population/property~\cite{Stromberg2007}. In order to account for large variance of damage, we devise a simple model by combining the occurrence distribution of natural disaster with population/property distributions, while the vulnerability is assumed to be constant. We will discuss each of three components in more detail.

Firstly, for population/property distributions, we consider two characteristics: probability density function (PDF) and spatial correlation. Some PDFs of population/property, denoted by $v$, are known to show power-laws as $P(v)\sim v^{-\alpha}$ with exponent $\alpha$. The estimated value of $\alpha$ for the wealth of the world's richest people over 1996--2012 ranges from $2$ to $3$~\cite{Brzezinski2013}, which can be related to Pareto principle~\cite{Newman2005}. The population distribution of cities in the United States follows a power-law with exponent $\alpha\approx 2.37$~\cite{Clauset2009}, consistent with Zipf's law. In addition, we consider spatial correlations of population/property because the spatial correlation can increase the variance of damage as exposed population/property are spatially concentrated due to the urbanization, such as Manhattan in New York City and Gangnam in Seoul. Areas with better accessibility may create more value, and the rents and infrastructural value of the areas could be higher. Company headquarters are likely to be located in such areas. It is also likely that the neighborhood of the rich (the poor) is rich (poor). Based on these observations we assume that the PDFs of population/property are characterized by power-law distributions as $P(v)\sim v^{-\alpha}$, and that those population/property are spatially correlated.

Secondly, the nature of natural disaster is considered. It is known that the intensity of natural disaster like earthquake, storm, and forest fire follows a power-law distribution~\cite{Nishenko1995,Bak2002,Clauset2009}. In our work, we focus on disasters like tornadoes that move along a trajectory. Since the intensity of disaster can be incorporated in modeling the vulnerability, we instead assume the length of a trajectory, denoted by $l$, to be distributed as power-law, $P(l)\sim l^{-\beta}$ with exponent $\beta$. For simplicity, we assume that each disaster is initiated at a random position, and moves in a random direction along a straight line of length $l$. The assumption of straight line can be easily relaxed to consider curved trajectories or even more complicated geometry. The assumption that the initiation position is not correlated with spatial configurations of population/property seems to be strong. In reality, people can choose to live in a location with less disasters to avoid damage. In contrast, people can prefer a location with more disasters if natural phenomena related to a certain disaster can benefit people despite possible damage by such disaster. For example, coastal area is more likely to be affected by tsunamis, while it provides ports for trade and fishing.  

Thirdly, we consider the vulnerability as a fraction of the realized damage out of each unit of population/property. It differs by variables such as wealth, building code, and network structure of infrastructure. Since there are more hospitals and more labor who are devoted to control disasters in cities, cities could have less vulnerability. On the other hand, cities could be more vulnerable due to a cascading effect of damage. In our model, since the property of vulnerability is hard to measure, we assume that the vulnerability is constant through the trajectory of disaster. Our model with the assumption of constant vulnerability can provide benchmark results for further realistic refinements. 

Finally, the total damage by a natural disaster is modeled to be as the sum of population/property exposed to that disaster, multiplied by the vulnerability of those population/property.

\begin{figure}[!b]
\centering\includegraphics[width=\columnwidth]{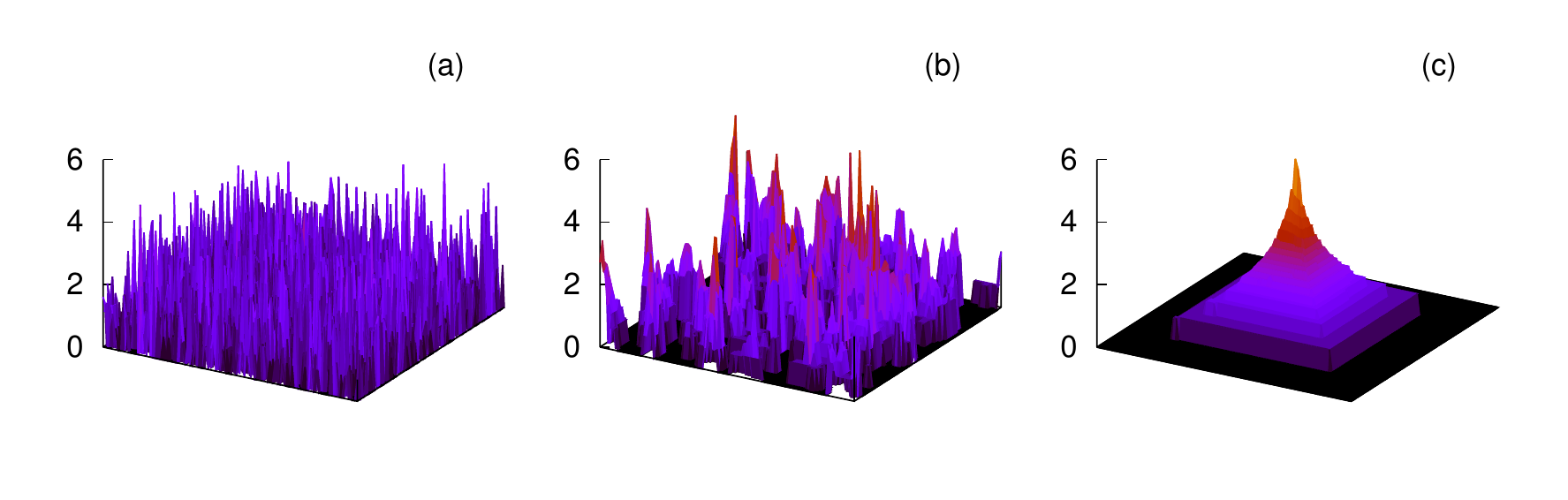}
\caption{(a) Random, (b) correlated, and (c) concentrated configurations of values on a two-dimensional lattice of size $100\times 100$, where the probability density function of value $v$ follows a power-law as $P(v)\sim v^{-\alpha}$ with $\alpha=2.5$. The height at each site represents a logarithm of the value.}
\label{fig:landscape}
\end{figure}

\subsection{Model setting} 

We first generate landscapes or configurations of population/property on a two-dimensional square lattice of size $L\times L$ with a periodic boundary condition, see Fig.~\ref{fig:landscape}. The population/property, or a value for convenience, at site $(i,j)$ is denoted by $v_{i,j}$ for $i,j=-\frac{L}{2},\cdots,\frac{L}{2}-1$. The PDF of value is assumed to follow a power law, $P(v)\sim v^{-\alpha}$ with exponent $\alpha>1$. To parameterize the degree of spatial correlation of value, we define a normalized centrality $c$ as a function of value configuration $\{v\}$:
\begin{eqnarray}
  \nonumber
  c(\{v\})&=&\frac {E_{\rm rand}-E(\{v\})}{E_{\rm rand} - E_{\rm conc}},\\
  E(\{v\})&=&\sum_{i,j}\left( \left|\ln \frac{v_{i,j}}{v_{i+1,j}}\right|+ \left|\ln \frac{v_{i,j}}{v_{i,j+1}}\right| \right),
  \nonumber
\end{eqnarray}
where $E$ measures the total difference between values of neighboring sites. $E_{\rm rand}$ and $E_{\rm conc}$ denote the values of $E$ for random and concentrated configurations, respectively. The zero centrality, $c=0$, corresponds to the random configuration, while the maximum centrality, $c=1$, implies that the values are concentrated in the central area, i.e., around the origin $(0,0)$. The configuration with intermediate $c$ is formulated using a simulated annealing algorithm. Starting from a random configuration, two randomly selected sites swap their values only if the swapping increases the correlation. The swapping is repeated until the correlation reaches the desired value of $c$. Figure~\ref{fig:landscape} shows exemplary configurations of value for random ($c=0$), correlated ($c\approx 0.85$), and concentrated ($c=1$) cases.

For natural disasters, we focus on moving disasters like tornadoes that move along a trajectory. We assume that a disaster initiated at a random site moves in a random direction, i.e., one of $\pm x$ and $\pm y$ directions, over the trajectory with length $l$. The length $l$ is randomly drawn from a distribution $P(l)\sim l^{-\beta}$ with exponent $\beta>1$. The vulnerability $A_{i,j}$ at site $(i,j)$ is assumed to be constant for all sites in the system such that $A_{i,j}=1$ for all $(i,j)$ for convenience. Then, the damage $D$ by the disaster initiated at $(i_0,j_0)$ and moving $l$ sites, say in the direction of $+x$-axis, is given as the sum of values over the trajectory: 
\begin{equation}
  \nonumber
  D(i_0,j_0,l)= \sum_{i=i_0}^{i_0+l-1} A_{i,j_0}v_{i,j_0} =\sum_{i=i_0}^{i_0+l-1} v_{i,j_0}.
\end{equation}

\section{Result}

It is expected that the damage $D$ has a large variance by showing a fat-tailed distribution, $P(D)\sim D^{-\gamma}$ with exponent $\gamma$. In general, the value of $\gamma$ depends on exponents $\alpha$, $\beta$, and the centrality $c$. 

\subsection{Random configurations}

\begin{figure}[!t]
\centering\includegraphics[width=\columnwidth]{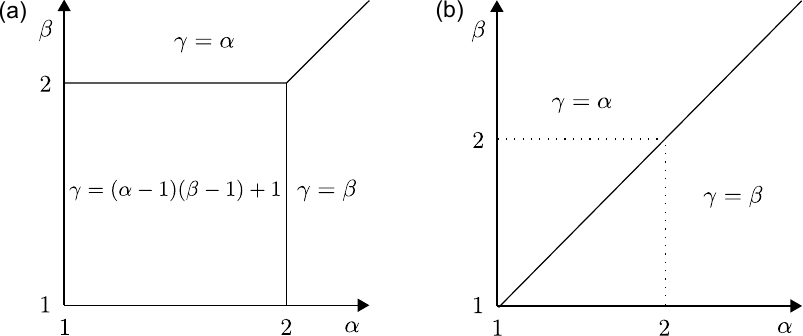}
\caption{Phase diagrams summarizing analytic results (a) for random configurations and (b) for concentrated configurations.}
\label{fig:phaseDiagram}
\end{figure}

\begin{figure*}[!t]
\centering\includegraphics[width=1.5\columnwidth]{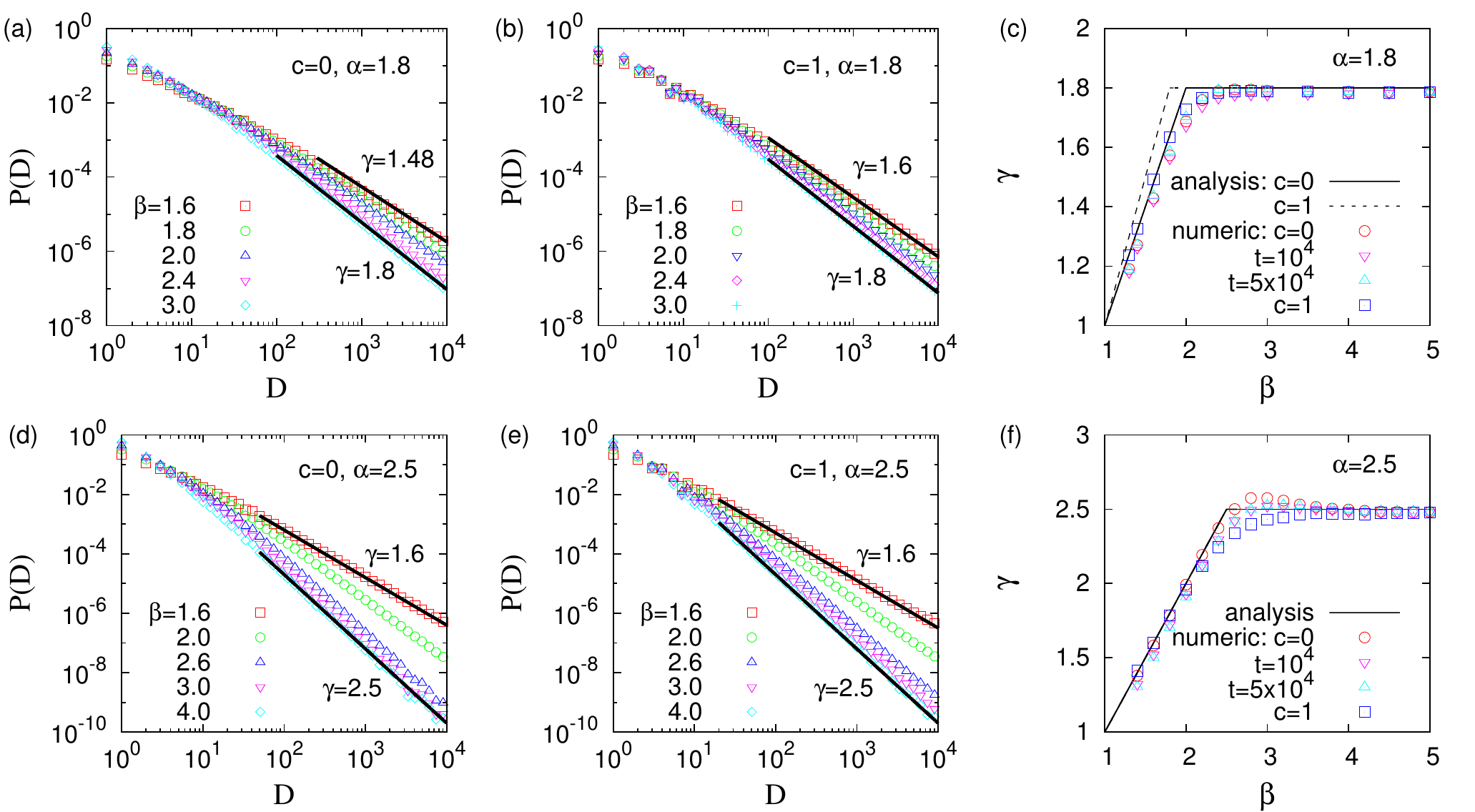}
\caption{Numerical results of damage distributions and their power-law exponents for $\alpha=1.8$ (top) and for $\alpha=2.5$ (bottom). In (c) and (f), $t$ denotes the Monte Carlo time in the simulated annealing to generate correlated configurations.}
\label{fig:numeric}
\end{figure*}

The case of random configurations with zero centrality can be analytically solved due to its uncorrelated nature~\cite{Jo2013}. The damage $D$ is independent of the initiation position and moving direction of the disaster, hence it can be written as a sum of $l$ independent and identical random variables, $v$s:
\begin{equation}
  \nonumber
  D=\sum_{n=1}^l v_n.
\end{equation}
For small $l$, as $l$ is mostly $1$, i.e., $D=v_1$, we obtain $D^{-\alpha}$ for $P(D)$. For sufficiently large $l$, if the variance of $\{v_n\}$ is small, one can approximate as $D\approx l\langle v\rangle$, where $\langle \cdot\rangle$ denotes an average, leading to $D^{-\beta}$ for $P(D)$. Finally, for sufficiently large $l$, if the variance of $\{v_n\}$ is large, $D$ is dominated by ${\rm max} \{v_n\}$ that is proportional to $l^{1/(\alpha-1)}$. By means of the identity $P(D)dD=P(l)dl$, one gets $D^{-(\alpha-1)(\beta-1)-1}$ for $P(D)$. We obtain apart from the coefficients
\begin{eqnarray}
  \nonumber
  P(D)&\sim& D^{-\alpha}+D^{-\beta}+D^{-(\alpha-1)(\beta-1)-1}\\
  \nonumber
  &\sim& D^{-\gamma},
\end{eqnarray}
thus for large $D$,
\begin{equation}
  \label{eq:gamma_c0}
\gamma=\min\{(\alpha-1)(\beta-1)+1,\alpha,\beta\},
\end{equation}
which is depicted in Fig.~\ref{fig:phaseDiagram}(a). This solution has been also obtained by rigorous calculations~\cite{Jo2013}. In case with $\alpha>2$ and $\alpha>\beta$, i.e., when the tail of value distribution is sufficiently thin, one obtains $\gamma=\beta$, implying that statistical properties of damage are determined only by those of disaster. In case with $\beta>2$ and $\beta>\alpha$, one gets $\gamma=\alpha$, implying the dominance of statistical properties of value in deciding damage. Only when both value and disaster distributions have sufficiently fat tails, i.e., when $\alpha,\beta<2$, the fat tail of damage can be explained in terms of the interplay of both value and disaster. We perform numerical simulations on the square lattice of linear size $L=3\cdot 10^3$ to confirm our analysis as shown in Fig.~\ref{fig:numeric}. 

\subsection{Concentrated configurations}

Since a concentrated configuration with $c=1$ has a rotational symmetry around the origin $(0,0)$, it can be described simply by a function of the distance $r$ from the origin, i.e., $v(r)\simeq v_0r^{-\mu}$ with $\mu=\frac{2}{\alpha-1}$. The relation $\mu=\frac{2}{\alpha-1}$ has been obtained by the identity $P(v)dv\propto 2\pi rdr$. For convenience, we calculate $D$ in a continuum limit of lattice as
\begin{eqnarray}
  \nonumber
  & &D(r_0,\theta_0,l,\phi) \simeq wv_0\int_0^l r(t)^{-\mu}dt,\\
  \nonumber
  & &r(t)=\sqrt{(r_0\cos\theta_0+t\cos\phi)^2+(r_0\sin\theta_0+t\sin\phi)^2},
\end{eqnarray}
where the polar coordinate $(r_0,\theta_0)$ and the angle $\phi$ are the initiation position and moving direction of the disaster, and $w$ denotes the transverse dimension or width of the disaster. Since $r(t)$ can be written in terms of $r_x\equiv r_0\cos(\phi-\theta_0)$ and $r_y\equiv r_0\sin(\phi-\theta_0)$, we get
\begin{eqnarray}
  \label{eq:D_c1}
  D\simeq wv_0\int_0^l[(t+r_x)^2+r_y^2]^{-\mu/2}dt.
\end{eqnarray}

For small $l$, the integration is approximated up to the first order of $l$, leading to $D\propto v_0 r_0^{-\mu}l\simeq v(r_0)l$. Thus, we obtain $D^{-\alpha}+D^{-\beta}$ for $P(D)$ apart from the coefficients. For large $l$, by substituting the variable of integration as $t+r_x=r_y\tan\theta$, one gets
\begin{eqnarray}
  \nonumber
  D\simeq wv_0r_y^{1-\mu}\int_{\tan^{-1}\frac{r_x}{r_y}}^{\tan^{-1}\frac{l+r_x}{r_y}}
\cos^{\mu-2}\theta d\theta.
\end{eqnarray}
Note that $\tan^{-1}\frac{r_x}{r_y}=\frac{\pi}{2}-(\phi-\theta_0)$. The following formula can be used:
\begin{eqnarray}
  \nonumber
  & &\int \cos^{\mu-2}\theta d\theta=\\
  & &\frac{\textrm{sgn}(\sin\theta)\cos^{\mu-1}\theta\ _2F_1(\frac{1}{2},\frac{\mu-1}{2};\frac{\mu+1}{2};\cos^2\theta)}{1-\mu}+\textrm{const.},
  \nonumber
\end{eqnarray}
where $\textrm{sgn}(x)$ gives the sign of $x$ and $_2F_1$ is the hypergeometric function. We consider two cases according to the moving direction of the disaster. In case with $|\phi-\theta_0|<\pi/2$, i.e., $r_x>0$, the disaster moves away from the central area. We get the result up to the leading terms as
\begin{eqnarray}
  \label{eq:D_c1_case1}
  D \approx wv_0\frac{l^{1-\mu}-a_1r_0^{1-\mu}}{1-\mu}
\end{eqnarray}
with a constant $a_1\equiv\ _2F_1(\frac{1}{2},\frac{\mu-1}{2};\frac{\mu+1}{2};\sin^2(\phi-\theta_0))$. If $\mu>1$ ($\alpha<3$), from $D\sim r_0^{1-\mu}\sim v^{\frac{\mu-1}{\mu}}$, we have the term $D^{-\frac{\alpha+1}{3-\alpha}}$ for $P(D)$. This is dominated by $D^{-\alpha}$ because $\frac{\alpha+1}{3-\alpha}>\alpha$ for $\alpha<3$. If $\mu<1$ ($\alpha>3$), $D\sim l^{1-\mu}$ leads to the term $D^{-\frac{(\alpha-1)(\beta-1)}{\alpha-3}-1}$ for $P(D)$, which is dominated by $D^{-\beta}$ for $\beta>1$. On the other hand, for $|\phi-\theta_0|>\pi/2$, i.e., $r_x<0$, the disaster approaches the central area to some extent and eventually moves away. The domain of integration in Eq.~(\ref{eq:D_c1}) can be divided into two at the closest position of the disaster to the origin given by $t_\times\equiv -r_x$: 
\begin{eqnarray}
  \nonumber
  \int_{t_\times}^l dt \leq \int_0^ldt=\int_{0}^{t_\times}dt + \int_{t_\times}^l dt \leq 2 \int_{t_\times}^l dt. 
\end{eqnarray}
Here the second inequality holds for sufficiently large $l$. Similarly to the case with $|\phi-\theta_0|<\pi/2$, we get the same result up to the leading terms as Eq.~(\ref{eq:D_c1_case1}) but with $a_1$ replaced by $a_2\equiv \sin^{1-\mu}(\phi-\theta_0)\Gamma(\frac{\mu+1}{2})\Gamma(\frac{1}{2})/\Gamma(\frac{\mu}{2})$. Finally, since $P(D)\sim D^{-\alpha}+D^{-\beta}$, we obtain the result for $\gamma$ as
\begin{equation}
  \label{eq:gamma_c1}
\gamma=\min\{\alpha,\beta\}.
\end{equation}
This solution is depicted in Fig.~\ref{fig:phaseDiagram}(b), and confirmed by numerical simulations as shown in Fig.~\ref{fig:numeric}.

\subsection{Correlated configurations}

Before investigating the effect of correlated configurations with $0<c<1$, we compare the results for random and concentrated cases, Eqs.~(\ref{eq:gamma_c0},~\ref{eq:gamma_c1}). If $\alpha>2$ or $\beta>2$, we get $\gamma=\min\{\alpha,\beta\}$ from $P(D)\sim D^{-\alpha}+D^{-\beta}$ for both cases of $c=0$ and $c=1$. The first term $D^{-\alpha}$ is mainly due to $D=v$ when $l=1$, hence it is independent of the spatial correlation or centrality $c$. The second term $D^{-\beta}$ is due to $D=\sum_{n=1}^l v_n\propto l$. That is, most disasters move along trajectories consisting of small $v_n$s when the tail of $P(v)$ is sufficiently thin, i.e., when $\alpha>2$. This leads to the irrelevance of the spatial correlation. Thus, one can expect that $\gamma=\min\{\alpha,\beta\}$ holds for the entire range of $c$. This is confirmed by numerical simulations for the case of $\alpha=2.5$ in Fig.~\ref{fig:numeric}(f), with some deviations mainly due to logarithmic corrections to scaling, like $\ln D$, and finite size effects. It is observed that the estimated values of $\gamma$ are systematically smaller for larger centrality, implying fatter tails of damage distributions.

For $\alpha<2$ and $\beta<2$, the difference in values of $\gamma$ for $c=0$ and for $c=1$ is summarized as follows:
\begin{eqnarray}
  \nonumber
  \Delta \gamma\equiv \gamma_{c=1}-\gamma_{c=0}=\left\{ \begin{tabular}{ll}
    $(2-\alpha)(\beta-1)$ & if $\beta<\alpha$,\\
    $(\alpha-1)(2-\beta)$ & if $\beta>\alpha$.
  \end{tabular}\right.
\end{eqnarray}
This implies that the tail of damage distribution for the concentrated case is always thinner than that for the random case. The maximum value of the difference $\Delta\gamma$ is $1/4$ when $\alpha=\beta=3/2$. The numerical simulations for the case of $\alpha=1.8$ in Fig.~\ref{fig:numeric}(c) confirm the analytic solution, with deviations due to corrections to scaling and finite size effects. While such deviations seem to be large, we systematically observe that in the region of $\beta<2$, the values of $\gamma$ for $c=1$ are slightly larger than those for $c=0$, comparable to the analytic results.

It turns out that whether the spatial correlation of value enhances or reduces the fat tail property of damage is not a simple issue as expected. The randomness in value configurations may enhance the variance of damage by introducing more fluctuations in exposed values when the tail of value distribution is sufficiently fat ($\alpha<2$). On the other hand, the randomness may reduce the variance of damage by mixing the values when the tail of value distribution is sufficiently thin ($\alpha>2$). The former explains the analytic expectation that the damage will have fatter tails for more correlated configurations, while the latter does the numerical observations of the opposite tendency.

\subsection{Empirical results}

\begin{figure}[!t]
\centering\includegraphics[width=\columnwidth]{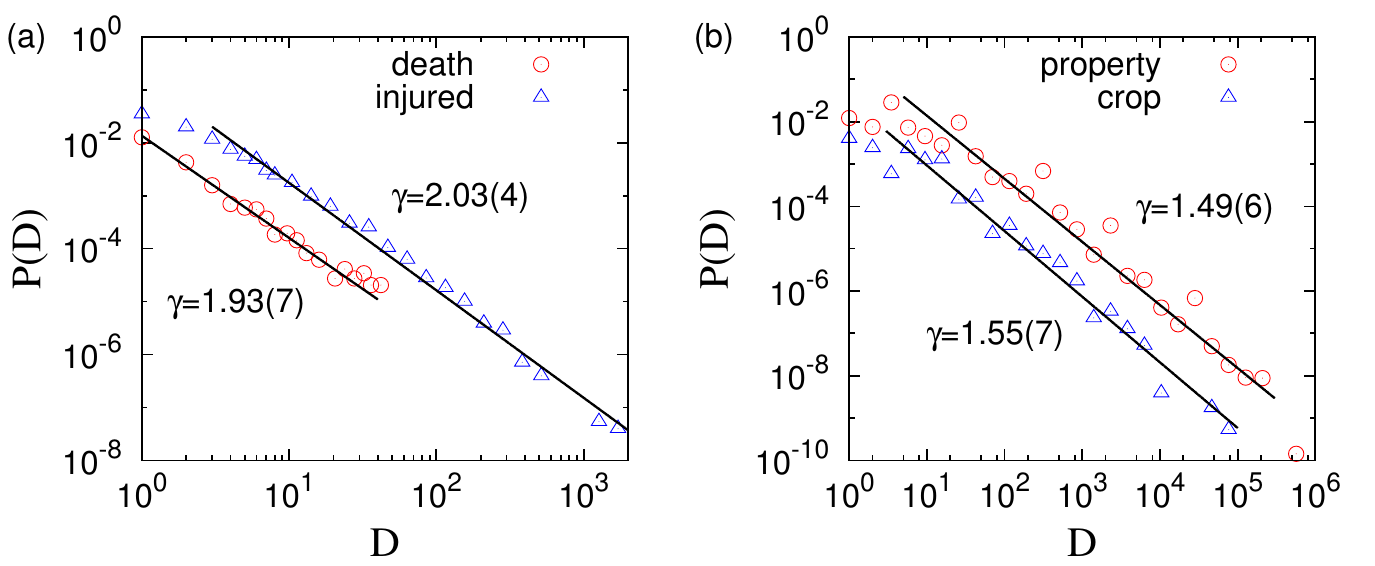}
\caption{Empirical damage distributions by tornadoes in the United States from 1970 to 2011 and estimated power-law exponents (a) for the numbers of death and injured and (b) for property and crop damages estimated in US dollars.}
\label{fig:tornado}
\end{figure}

In order to support our results, we empirically study casualty and property damage distributions by tornadoes in the United States from 1970 to 2011, for which the data were retrieved on 24 June 2011 from the website of National Climatic Data Center. By assuming a power-law form for those distributions, the power-law exponents are estimated as $\gamma\approx 2$ for the numbers of death and the injured, and $\gamma\approx 1.5$ for property and crop damages, as shown in Fig.~\ref{fig:tornado}. The significantly different values of power-law exponent, i.e., $1.5$ and $2$, could imply that they represent qualitatively different underlying mechanisms or origins. 

In order to account for these observations for tornadoes, our simple model can be extended to take into account various factors like position-dependent vulnerability. For example, let us consider that the vulnerability at a site scales with the value at that site as $A\sim v^\eta$, where the scaling exponent $\eta$ can be positive or negative depending on the situation. Since the effective value, denoted by $v'\equiv Av$, is proportional to $v^{1+\eta}$, the fat tail of the PDF of effective value is characterized by the power-law exponent $\alpha'=\frac{\alpha+\eta}{1+\eta}$. Note that $\alpha'$ reduces to $\alpha$ for $\eta=0$ as in our simplest setup. More detailed analysis for the effect of position-dependent vulnerability is left for future works.

\section{Conclusion}

We have developed a simple model to show that damages by natural disasters could have large variances in terms of fat-tailed distributions of natural disaster and population/property, as well as in terms of their spatial correlations. The damage has been modeled as the sum of population/property exposed to the moving disaster, while the vulnerability was assumed to be constant through the trajectory of disaster. Our simple model draws limits to the implication of the results. In reality, vulnerability differs by variables such as wealth, building code, and network structure of infrastructure. The trajectory of disaster may be not straight. However, our model can still provide the benchmark results for more realistic refinements, for which these assumptions can be easily relaxed. 

Our research enables to quantitatively study the effect of fat tail property, in terms of the exact analytic results for the power-law exponent of damage distributions. Thus, our model can serve as more concrete framework for future studies on damage by natural disaster as well as risk analysis under uncertainty with fat-tailed distributions. We also note that since a portion of damages due to climate change is associated with natural disaster, our research can provide grounds to the discussion on the fat tail property of damage due to climate change~\cite{Weitzman2011}.

\begin{acknowledgements}
Financial support by the Aalto University postdoctoral programme (H.-H.J.) is gratefully acknowledged.
\end{acknowledgements}


\begin{thebibliography}{12}
\expandafter\ifx\csname natexlab\endcsname\relax\def\natexlab#1{#1}\fi
\expandafter\ifx\csname bibnamefont\endcsname\relax
  \def\bibnamefont#1{#1}\fi
\expandafter\ifx\csname bibfnamefont\endcsname\relax
  \def\bibfnamefont#1{#1}\fi
\expandafter\ifx\csname citenamefont\endcsname\relax
  \def\citenamefont#1{#1}\fi
\expandafter\ifx\csname url\endcsname\relax
  \def\url#1{\texttt{#1}}\fi
\expandafter\ifx\csname urlprefix\endcsname\relax\def\urlprefix{URL }\fi
\providecommand{\bibinfo}[2]{#2}
\providecommand{\eprint}[2][]{\url{#2}}

\bibitem[{\citenamefont{Imamura and Anawat}(2012)}]{Imamura2012}
\bibinfo{author}{\bibfnamefont{F.}~\bibnamefont{Imamura}} \bibnamefont{and}
  \bibinfo{author}{\bibfnamefont{S.}~\bibnamefont{Anawat}}, in
  \emph{\bibinfo{booktitle}{Proceedings of the International Symposium on
  Engineering Lessons Learned from the 2011 Great East Japan Earthquake, March
  1-4, 2012, Tokyo, Japan}} (\bibinfo{year}{2012}), pp.
  \bibinfo{pages}{21--30}.

\bibitem[{\citenamefont{Nishenko and Barton}(1995)}]{Nishenko1995}
\bibinfo{author}{\bibfnamefont{S.~P.} \bibnamefont{Nishenko}} \bibnamefont{and}
  \bibinfo{author}{\bibfnamefont{C.}~\bibnamefont{Barton}},
  \emph{\bibinfo{title}{Scaling Laws for Natural Disaster Fatalities}}
  (\bibinfo{year}{1995}), U.S. Geological Survey Open File Report No. 95-67.

\bibitem[{\citenamefont{Bak et~al.}(2002)\citenamefont{Bak, Christensen, Danon,
  and Scanlon}}]{Bak2002}
\bibinfo{author}{\bibfnamefont{P.}~\bibnamefont{Bak}},
  \bibinfo{author}{\bibfnamefont{K.}~\bibnamefont{Christensen}},
  \bibinfo{author}{\bibfnamefont{L.}~\bibnamefont{Danon}}, \bibnamefont{and}
  \bibinfo{author}{\bibfnamefont{T.}~\bibnamefont{Scanlon}},
  \bibinfo{journal}{Physical Review Letters} \textbf{\bibinfo{volume}{88}},
  \bibinfo{pages}{178501+} (\bibinfo{year}{2002}).

\bibitem[{\citenamefont{Clauset et~al.}(2009)\citenamefont{Clauset, Shalizi,
  and Newman}}]{Clauset2009}
\bibinfo{author}{\bibfnamefont{A.}~\bibnamefont{Clauset}},
  \bibinfo{author}{\bibfnamefont{C.~R.} \bibnamefont{Shalizi}},
  \bibnamefont{and} \bibinfo{author}{\bibfnamefont{M.~E.~J.}
  \bibnamefont{Newman}}, \bibinfo{journal}{SIAM Review}
  \textbf{\bibinfo{volume}{51}}, \bibinfo{pages}{661} (\bibinfo{year}{2009}).

\bibitem[{\citenamefont{Krugman}(1996)}]{Krugman1996}
\bibinfo{author}{\bibfnamefont{P.~R.} \bibnamefont{Krugman}},
  \emph{\bibinfo{title}{The self-organizing economy}}
  (\bibinfo{publisher}{Blackwell Publishers}, \bibinfo{year}{1996}),
  \bibinfo{edition}{1st} ed.

\bibitem[{\citenamefont{Weitzman}(2009)}]{Weitzman2009}
\bibinfo{author}{\bibfnamefont{M.~L.} \bibnamefont{Weitzman}},
  \bibinfo{journal}{Review of Economics and Statistics}
  \textbf{\bibinfo{volume}{91}}, \bibinfo{pages}{1} (\bibinfo{year}{2009}).

\bibitem[{\citenamefont{Nordhaus}(2011)}]{Nordhaus2011}
\bibinfo{author}{\bibfnamefont{W.~D.} \bibnamefont{Nordhaus}},
  \bibinfo{journal}{Review of Environmental Economics and Policy}
  \textbf{\bibinfo{volume}{5}}, \bibinfo{pages}{240} (\bibinfo{year}{2011}).

\bibitem[{\citenamefont{Str\"{o}mberg}(2007)}]{Stromberg2007}
\bibinfo{author}{\bibfnamefont{D.}~\bibnamefont{Str\"{o}mberg}},
  \bibinfo{journal}{The Journal of Economic Perspectives}
  \textbf{\bibinfo{volume}{21}}, \bibinfo{pages}{199} (\bibinfo{year}{2007}).

\bibitem[{\citenamefont{Brzezinski}(2013)}]{Brzezinski2013}
\bibinfo{author}{\bibfnamefont{M.}~\bibnamefont{Brzezinski}},
  \emph{\bibinfo{title}{Do wealth distributions follow power laws? evidence
  from ``rich lists"}} (\bibinfo{year}{2013}),
  \urlprefix\url{http://arxiv.org/abs/1304.0212}.

\bibitem[{\citenamefont{Newman}(2005)}]{Newman2005}
\bibinfo{author}{\bibfnamefont{M.~E.~J.} \bibnamefont{Newman}},
  \bibinfo{journal}{Contemporary Physics} \textbf{\bibinfo{volume}{46}},
  \bibinfo{pages}{323} (\bibinfo{year}{2005}).

\bibitem[{\citenamefont{Jo et~al.}(2013)\citenamefont{Jo, Pan, Perotti, and
  Kaski}}]{Jo2013}
\bibinfo{author}{\bibfnamefont{H.-H.} \bibnamefont{Jo}},
  \bibinfo{author}{\bibfnamefont{R.~K.} \bibnamefont{Pan}},
  \bibinfo{author}{\bibfnamefont{J.~I.} \bibnamefont{Perotti}},
  \bibnamefont{and} \bibinfo{author}{\bibfnamefont{K.}~\bibnamefont{Kaski}},
  \bibinfo{journal}{Physical Review E} \textbf{\bibinfo{volume}{87}},
  \bibinfo{pages}{062131+} (\bibinfo{year}{2013}).

\bibitem[{\citenamefont{Weitzman}(2011)}]{Weitzman2011}
\bibinfo{author}{\bibfnamefont{M.~L.} \bibnamefont{Weitzman}},
  \bibinfo{journal}{Review of Environmental Economics and Policy}
  \textbf{\bibinfo{volume}{5}}, \bibinfo{pages}{275} (\bibinfo{year}{2011}).

\end{thebibliography}
\end{document}